\documentclass[10pt,a4paper]{article}

\usepackage[T1]{fontenc}
\usepackage[utf8]{inputenc}
\usepackage[american]{babel}

\usepackage{cite}
\usepackage{amsmath,amssymb,amsfonts}
\usepackage{algorithm}
\usepackage{algpseudocode}
\usepackage{graphicx}
\usepackage{textcomp}

\usepackage{multicol}

\usepackage{subcaption}

\usepackage{url}

\usepackage{color}
\usepackage{xcolor}

\begin{document}

\title{Detection of Silent Data Corruptions in \\Smoothed Particle Hydrodynamics Simulations}


\author{Aurélien Cavelan\\
\textit{University of Basel, Switzerland}\\
\textit{aurelien.cavelan@unibas.ch}
\and
Rubén M. Cabezón\\
\textit{University of Basel, Switzerland}\\
\textit{ruben.cabezon@unibas.ch}
\and
Florina M. Ciorba\\
\textit{University of Basel, Switzerland}\\
\textit{florina.ciorba@unibas.ch}
}

\maketitle

\begin{abstract}

Silent data corruptions~(SDCs) hinder the correctness of long-running scientific applications on large scale computing systems. Selective particle replication~(SPR) is proposed herein as the first particle-based replication method for detecting SDCs in Smoothed particle hydrodynamics~(SPH) simulations. SPH is a mesh-free Lagrangian method commonly used to perform hydrodynamical simulations in astrophysics and computational fluid dynamics. SPH performs interpolation of physical properties over neighboring discretization points (called SPH particles) that dynamically adapt their distribution to the mass density field of the fluid. When a fault (e.g., a bit-flip) strikes the computation or the data associated with a particle, the resulting error is silently propagated to all nearest neighbors through such interpolation steps. SPR replicates the computation and data of a few carefully selected SPH particles. SDCs are detected when the data of a particle differs, due to corruption, from its replicated counterpart. SPR is able to detect many DRAM SDCs as they propagate by ensuring that all particles have at least one neighbor that is replicated. The detection capabilities of SPR were assessed through a set of error-injection and detection experiments and the overhead of SPR was evaluated via a set of strong-scaling experiments conducted on an HPC system. The results show that SPR achieves detection rates of 91-99.9\%, no false-positives, at an overhead of 1-10\%.
\end{abstract}


\section{Introduction}

The frequency of faults, errors and failures in large-scale systems increases proportionally to the number of components. As such, reliability has
been identified as a major challenge for \mbox{Exascale} computing~\cite{Snir_FailuresExascale,IESP-toward, JSFI14}.
Trustworthy computing aims at guaranteeing the correctness of the results of \mbox{long-running} computations on \mbox{large-scale} supercomputers.
In particular, silent errors, or \emph{silent data corruptions} (SDCs), represent a major threat to the correctness of the results of such calculations. 
SDCs typically manifest when one or more bits are flipped in the memory of the system.
There are several causes for these errors, such as packaging pollution or cosmic rays, among others~\cite{BGomez:2016,Ogorman94,Ziegler98
}.

The standard solution for protecting the memory subsystem against data corruption is the use of error correcting codes (ECC). However, these mechanisms are not able to detect and correct all errors.
Indeed, recent studies suggest that such mechanisms may not prevent data corruptions from occurring at extreme scales~\cite{BGomez:2016,GPUerrors,Hwang:2012}, and in particular in DRAM devices~\cite{Sridharan:2015:MEM:2694344.2694348}.
This work expands on detecting DRAM errors.

Triple modular \emph{redundancy}\footnote{In this work, \emph{redundancy} and \emph{replication} are used interchangeably.} (TMR)~\cite{Lyons1962} is the most general and non-intrusive approach to detect and correct SDCs in scientific applications.
SDCs detectors are characterized by their precision and their recall.
The precision is the ratio of true errors detected (in contrast to false-positives) over all errors detected, and the recall is simply the ratio of true errors detected over all true errors that occurred during the execution.
A high precision and a high recall indicate both few false-positives and a good detection rate, respectively.
In general, detectors that either employ \emph{full replication} of entire applications~\cite{Fiala12Detection} or \emph{selective replication} of parts of an application~\cite{Berrocal17Toward} offer the highest precision and recall.
However, they are often prohibitively expensive in terms of additional required computing resources and time.
Therefore, many application-specific detectors have been proposed as alternatives to redundancy to lower the cost of error detection. 
Specifically, \emph{data analytics-}based techniques detect outliers by relying on application-specific properties, such as spatial and/or temporal data smoothness. 
\emph{Interpolation-}based detectors employ techniques such as time series prediction and spatial multivariate interpolation~\cite{PPoPP14, Gomez15Detecting, Gomez15Exploiting} to interpolate the next value of a data-point, and offer broad error detection coverage at low cost, but typically have a lower precision compared to full-replication approaches.
A further review of relevant error detection methods is presented in Section~\ref{sec:relwork}.

\emph{Smoothed particle hydrodynamics} (SPH) is a meshless \mbox{Lagrangian} method commonly used for performing hydrodynamical simulations.
Initially devised in the late '70s~\cite{lucy1977,gingold1977
}, this technique underwent sustained development ~\cite{cabezon2008,
springel2010,
garciasenz2012,
rosswog2015
}  
and is, nowadays, commonly used in many fields including computational fluid dynamics, plasma physics, solid mechanics, and astrophysics.
Its inherent good conservation properties and its adaptability to distorted geometries make SPH a common choice to simulate highly-dynamic three-dimensional scenarios.
The SPH technique discretizes a fluid in a series of interpolation points (called SPH particles) whose distribution follows the mass density of the fluid, and their evolution relies then on a weighted interpolation over close neighboring particles. The cost of such a highly-adaptive unstructured fluid discretization is the \emph{non-uniformity of data}, in terms of the number and distribution of neighbors that a particle might have at a specific location and time. This renders most \mbox{data~analytics-based} error detectors less accurate than redundancy-based detectors. Moreover, SPH simulations are inherently robust: it is likely that SDCs will not crash the execution of an SPH application, thereby not alerting the user.

\emph{\mbox{Selective} particle replication}~(SPR) is proposed in this work.
SPR replicates the computation and data of a few carefully selected SPH particles, and SDCs are detected when the data of a particle differs from the data of its replica or viceversa.

It is necessary to understand how SDCs propagate in SPH simulations before we clarify how to select which particles need to be replicated. 
We assume that an SDC initially strikes the data of a single particle (e.g., a bit-flip in the position, density, mass, temperature, or in any other data field associated with that particle).
The subsequent SPH interpolation steps will proceed with erroneous data and propagate the error to all neighbors of that particle, i.e., typically affecting $\sim100$ other particles.
However, having just one redundant neighbor would be enough to detect the difference (no matter how small) in the computed result due to the original SDC.
Therefore, by ensuring that \textit{all particles have at least one redundant neighbor}, SPR can detect most SDCs (as they propagate).
We detail the entire process and its limitations in Section~\ref{sec:SPR}.

We conduct a series of experiments by incorporating SPR in a production SPH code and running two test simulations commonly used in astrophysics.
First, we perform a set of error-injection and detection experiments to assess the detection capabilities of SPR.
Next, we conduct a set of performance-evaluation experiments, without injecting errors, to assess the overhead of SPR for the two test simulations on the Piz Daint supercomputer\footnote{\url{https://www.cscs.ch/computers/piz-daint/}}. 
The outcome shows that SPR is particularly suited for detecting SDCs in SPH simulations, detecting $91-99.9\%$ of the injected errors, for a typical execution overhead between $1-10\%$. In addition, SPR raises no false-positives, which is a significant advantage over unprecise detectors.

This work makes the following contributions:
(1)~Presents a simple, yet effective method (SPR) for selecting particles to be replicated and detects errors with bit-level precision;
(2)~Experimentally quantifies the recall and precision of SPR through an error-injection and detection campaign in a production scientific application; and 
(3)~Assesses the scalability and overhead of the proposed SPR approach through real experiments on an HPC system.

The remainder of this work is structured as follows.
We describe in detail and illustrate the proposed approach for silent error detection via \textit{selective particle replication} in Section~\ref{sec:SPR}.
We include the experimental setup and results in Section~\ref{sec:experiments} along with their evaluation and a discussion of the findings.
Section~\ref{sec:relwork} surveys the relevant literature related to silent errors detection. Finally, Section~\ref{sec:conclusion} outlines the conclusion of this work together with insights into future work extensions.

\section{Selective Particle Replication}\label{sec:SPR}

\emph{Selective particle replication}~(SPR) consists of three steps, namely \textit{selection}, \textit{replication}, and \textit{detection}.
Section~\ref{sec:algo-selection} describes an algorithm to select the particles that will be replicated, Section~\ref{sec:replication} covers the different aspects and challenges of particle replication in SPH, and Section~\ref{sec:algo-detection} presents the error-detection algorithm as well as its limitations.

\subsection{Selection}\label{sec:algo-selection}

\begin{figure*}[!htb]
	\centering
		\includegraphics[scale=0.85,angle=-90]{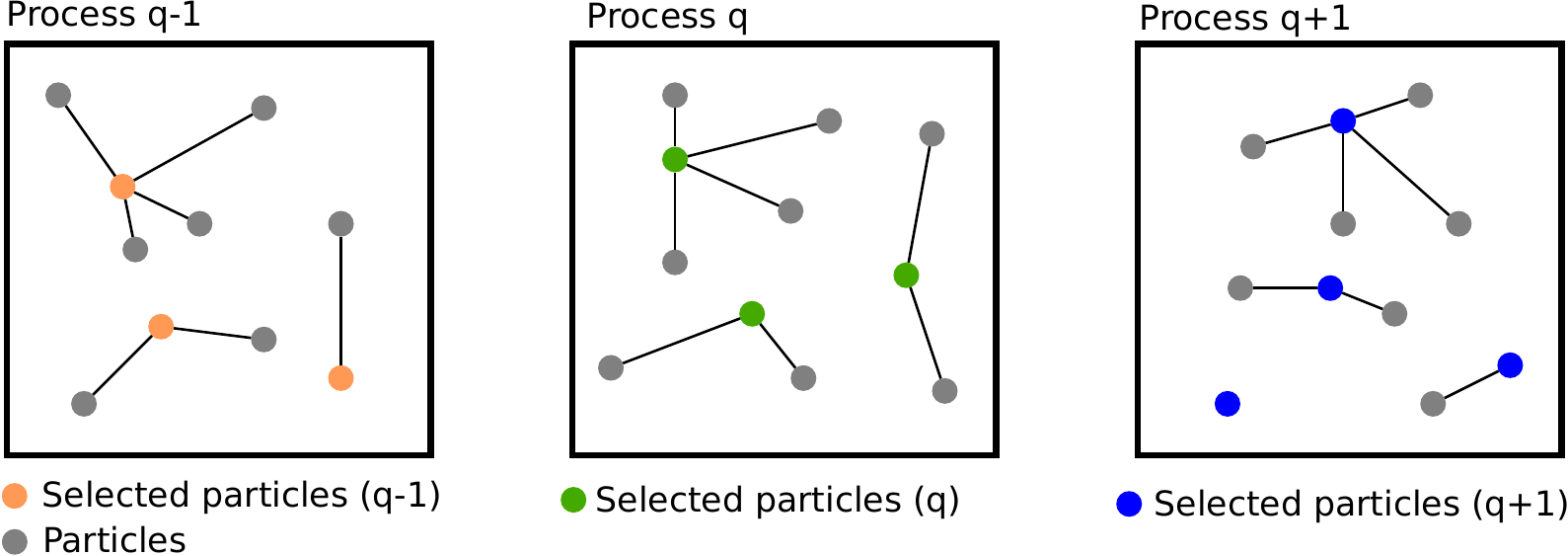}
	\caption{SPR step 1: selection of particles to replicate. The figure shows the distribution of $32$ particles on three processes $\{q-1,q,q+1\}$ computing on particles from different sections of the SPH domain. Selected particles on each process form a Maximal Independent Set (MIS) such that all particles have at least one neighbor that is selected for replication in another process. Edges show the connection between the selected particles and their neighbors.}
	\label{fig.SPR-selection}
\end{figure*}

In parallel SPH implementations~\cite{ParallelSPH}, the particle domain is distributed among the processes.
Each process is given a list of particles and their associated data.
The first step in SPR is to select which particles to replicate to ensure that all particles in a process have at least one neighbor that is replicated in another process, as illustrated in Figure~\ref{fig.SPR-selection}.


Given that the number of computations grows linearly with the number of particles selected on each process, we want to replicate as few particles as possible overall. The domain composed of the particles and their neighbors can be compared to a graph where vertices and edges represent the particles connected to their neighbors.

Particles must be selected such that any given particle in a process either belongs to the set of particles selected for replication, or has a selected particle as neighbor, which translates to finding a \emph{Minimum Dominating Set} (MDS). 
However, this is a classical NP-Hard problem~\cite{NP-Hard:1990} and there is no known optimal polynomial time algorithm for finding such a minimum set.
One can alternatively solve the simpler \emph{Maximal Independent Set} (MIS) problem, where a MIS is also a Dominating Set (albeit not minimum). 
As opposed to a dominating set, a MIS ensures that no selected vertices (respectively particles) in the graph are adjacent, which leads to fewer vertices being selected, and it is possible to find many MISs in a graph.


The present work proposes the use of a greedy algorithm to find a MIS, listed in Algorithm~\ref{fig.algo.selection}.
Given the set $P_q$ of particles assigned to process $q$ for computation, Algorithm~\ref{fig.algo.selection} returns the set $S_q$ of particles selected for replication in another process.
A possible solution to the problem is presented in Figure~\ref{fig.SPR-selection} for a small example with $32$ particles distributed on three processes, highlighting the selected particles in each process.

\begin{algorithm} 
\caption{Particle Selection} 
\label{fig.algo.selection}
\begin{algorithmic}
\Procedure{select-particle}{$P_q$}
  \State Initialize \textit{$S_q$} to an empty set
  \While{$P_q$ is not empty}
  	\State Choose a particle $p \in P_q$
  	\State Add $p$ to \textit{$S_q$}
  	\State Remove from $P$ the particle $p$ and all its neighbors
  \EndWhile
  \State \Return \textit{$S_q$}
\EndProcedure 
\end{algorithmic}
\end{algorithm}

\noindent\textbf{Complexity.} Algorithm~\ref{fig.algo.selection} requires, in the worst case, to iterate through the $n_q$ particles of process $q$ and for each particle to visit at most $ng_{max}$ neighbors, where $ng_{max}$ denotes the maximum number of neighbors for any given particle in the simulation. Therefore, the running time of this algorithm is $\mathcal{O}(n_q ng_{max})$. 

\noindent\textbf{Overall asymptotic cost.} The running time of SPH applications is dominated by large interpolation kernels scaling as $O(n_q ng_{max})$. Comparatively, our approach only requires $\mathcal{O}(n_r ng_{max})$ additional computations, where $n_r$ is the maximum number of replicated particles per node. A lower bound on $n_r$ is $\frac{1}{ng_{max}}$, when only $1$ particle is selected to dominate all of its neighbors. Experiments show $n_r$ to be in the range of $1-10\%$ based on our MIS approach. Consequently, the increment in memory cost and communication time is also very small $1-10\%$ (see Section~\ref{sec:experiments}).

Note that the maximum number of neighbors, $ng_{max}$, is often a user-defined parameter in the order of $10^2$~particles for 3D simulations.\footnote{Given a target number of neighbors, the simulation will try to reach this number of neighbors for each particle and this influences the resulting smoothing length.}
As the simulated scenarios grow in complexity and require increased resolution, the overall number of particles in SPH simulations increases, as well as the number of neighbors. 
Nowadays, 3D SPH simulations require between $10^5 - 10^{12}$~particles, with $10^2 - 5 \cdot 10^2$ neighbors per particle.

\subsection{Replication}\label{sec:replication}

\begin{figure*}[!htb]
	\centering
		\includegraphics[scale=0.85,angle=-90]{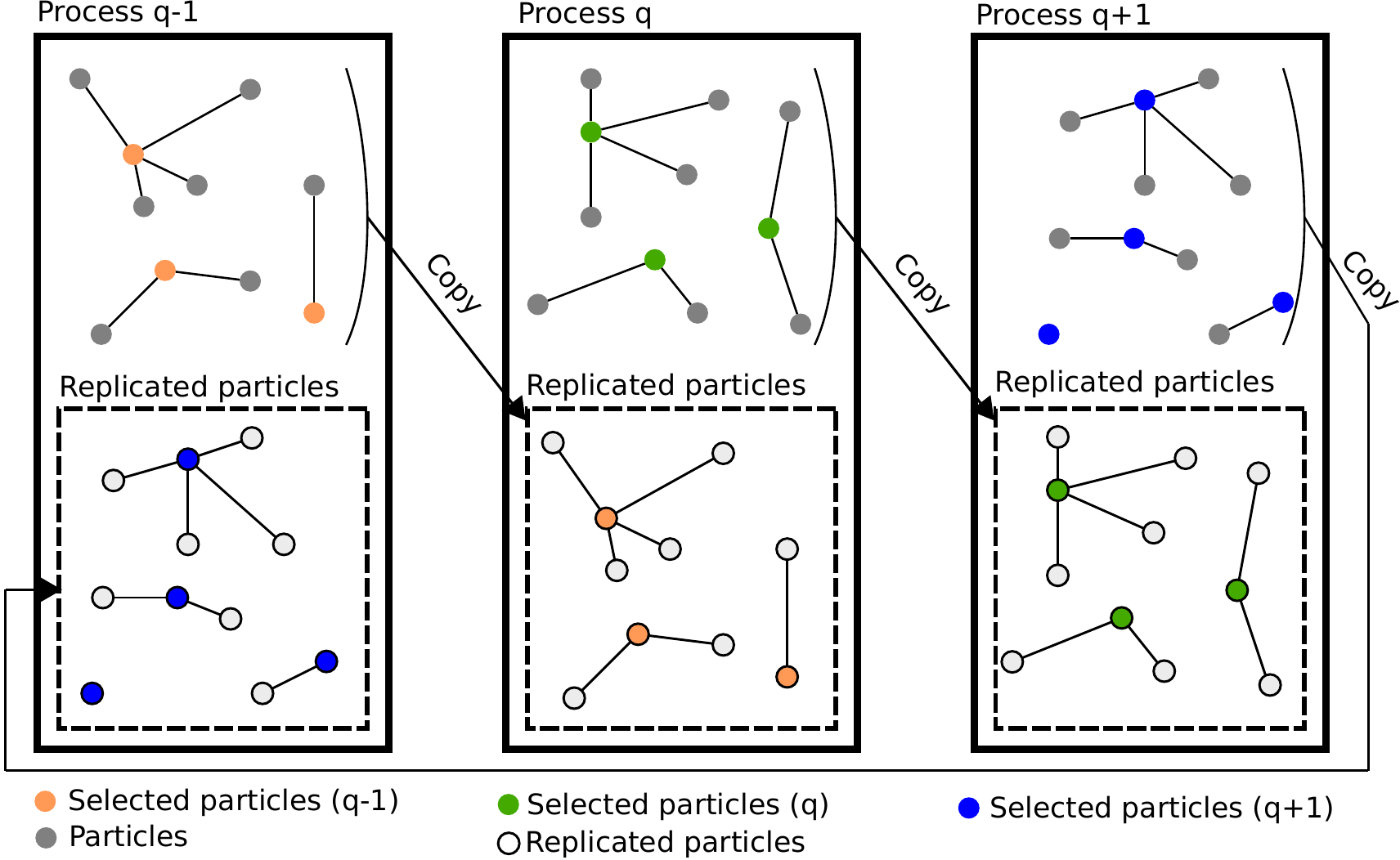}
	\caption{SPR step 2: replication of selected particles onto the target process. Each process sends a copy of the data associated with all its particles to the next process id. Note that all particles are copied, but only selected particles (orange, green, blue) are computed upon the target process, using the data from their replicated neighbors (light gray).}
	\label{fig.SPR-replication}
\end{figure*}

The second SPR step is to copy the data and to replicate the execution of the previously selected particles onto a target process.
Each process must send a copy of all its selected particles and their neighbors to its assigned target process.
As the selected particles are in fact a MIS, each process copies of the data of \textit{all} its particles on to its assigned target process, as shown in Figure~\ref{fig.SPR-replication}. However, only the selected particles are computed on the target process.

Figure~\ref{fig.SPR-replication} shows the distribution of $32$ particles onto three processes $\{q-1,q,q+1\}$.
In this configuration, each process initially sends a copy of the data associated with all its particles to the next process id.
Note that both computation and data are replicated onto a target process for the selected particles, while only data is replicated for all the neighbors of a selected particle. This data is needed to update the selected particles during SPH neighbor interpolation steps.

To avoid the risk of using corrupted data, detection must be performed before any inter-process communications occur. As long as no data is communicated to another process, the error is guaranteed to be contained within the faulty process. Since in SPH, processes must exchange data after every interpolation step to exchange neighbors that are at the edge of their respective computational domain, SPR detection must also be done right after each interpolation step.
In addition, each process must update the copy of the data of all its particles on the target process, so that the selected replicated particles use the latest version of their neighbors data. 

\begin{algorithm}
\caption{SPH General Computational Workflow with SPR}
\label{fig.algo.sph+gr}
\begin{algorithmic}
 \State Initialization
 \While{Target simulated time is not reached}
 	\State 1. Build tree
 	\State \quad1.1 \textcolor{blue}{Detection}
 	\State \quad1.2 \textcolor{blue}{Update of neighboring particles}
 	\State 2. Find neighbors
 	\State \quad2.1 \textcolor{blue}{Detection}
 	\State \quad2.2 \textcolor{blue}{Select particles for replication}
 	\State \quad2.3 \textcolor{blue}{Update of neighboring particles}
 	\State 3. Execute SPH neighboring interpolation kernels
 	\State \quad3.1 \textcolor{blue}{Detection}
 	\State \quad3.2 \textcolor{blue}{Update of neighboring particles}
 	\State 4. Find new time-step
 	\State \quad4.1 \textcolor{blue}{Detection}
 	\State \quad4.2 \textcolor{blue}{Update of neighboring particles}
 	\State 5. Update velocity and position
 	\State \quad5.1 \textcolor{blue}{Detection}
 	\State \quad5.2 \textcolor{blue}{Update of neighboring particles}
 	\State 6. (Optional) Compute self-gravity
 	\State \quad6.1 \textcolor{blue}{Detection}
 	\State \quad6.2 \textcolor{blue}{Update of neighboring particles}
 \EndWhile
\end{algorithmic}
\end{algorithm}

The general workflow is described in Algorithm~\ref{fig.algo.sph+gr} and additional SPR steps are shown in blue.
Depending on the scenario they simulate or the research field in which they are used, SPH codes can greatly vary in terms of implementation and physical processes they include. 
Nevertheless, many SPH codes apply the same general underlying workflow. 
This allows us to generalize the use of the SPR method proposed here, not only to a single specific SPH code but to the vast majority of SPH codes.

\subsection{Detection}\label{sec:algo-detection}

Error-detection is done by comparing the data of the selected particles against their the data of their replica.
Detection must be performed after every computational workflow step, before any data are communicated to other processes, as illustrated in Algorithm~\ref{fig.algo.sph+gr} (in blue).

Given that an original particle and its replica represent the same particle, their data are expected to remain identical at any point during the simulation. 
Moreover, the data of the original particle can be compared against the data of its replica with \textit{bit-level resolution}. 
If an error propagates to either the original particle or its replica, even by causing a change in single bit, it will be detected via a simple comparison.

\begin{figure*}[!htb]
	\centering
		\includegraphics[scale=0.85,angle=-90]{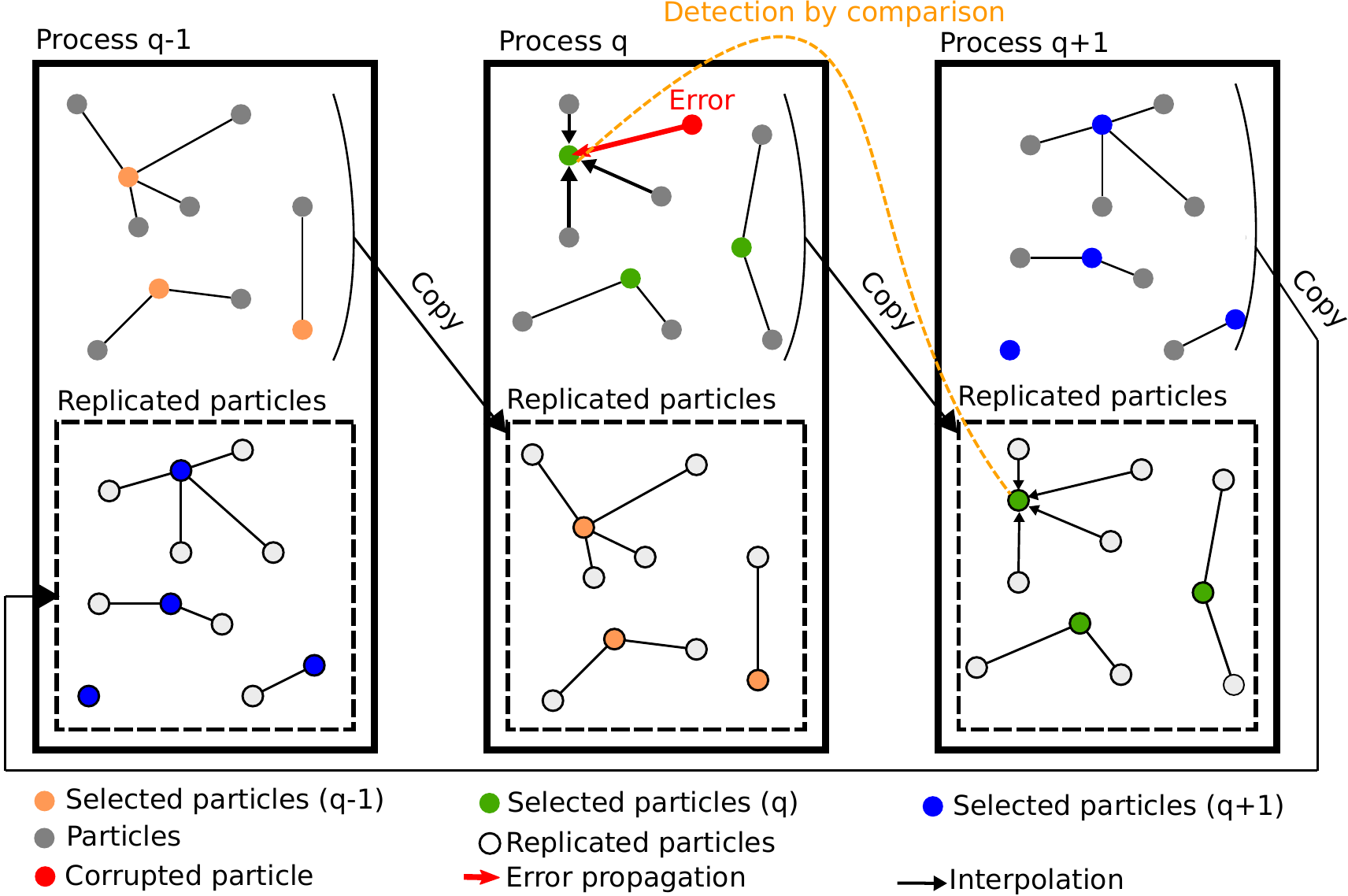}
	\caption{SPR step 3: error detection. An error (e.g. a bit flip) has corrupted the data associated with a particle (in red). When the selected particle (in green) that is a neighbor of the corrupted particle is updated via interpolation over its neighbors, the error is propagated and the data of the selected particle becomes corrupted. SPR detects the effects of this silent error from the corrupted particle by comparing the data of the selected particles on $q$ with their replica on the target process $q+1$.}
	\label{fig.SPR-detection}
\end{figure*}

Figure~\ref{fig.SPR-detection} shows how one of the selected particle on process~$q$ becomes corrupted after using data from a corrupted neighbor during an interpolation step. Since the newly computed data on process $q$ differs from the data of its replica on process $q+1$, the error is detected via a simple comparison.

The following algorithm is proposed to detect errors using replicated particles selected \textit{a priori} in Step 1.
We assume that replicated particles are always computed upon the next process in a round robin fashion, i.e., the list of replicated particles selected on process $q$ with Algorithm~\ref{fig.algo.selection} is sent to process $q+1$, so that process $q+1$ computes the selected replicated particles from process $q$, and so on.
Therefore, for detection purposes, process $q$ sends the data corresponding to replicated particles from process $q-1$ back to process $q-1$, and in return, it receives the data corresponding to its own replicated particles from process $q+1$.
Let $S_{q-1}$ denote the set of particles selected for replication on process $q-1$ and let $data_q(S_{q-1})$ denote the data associated with replicated particles selected on process $q-1$ and used for computation on process $q$. Similarly, let $data_{q+1}(S_{q})$ denote the data associated with replicated particles selected on process $q$ and used for computation on process $q+1$.
The detection algorithm is listed in Algorithm~\ref{fig.algo.detection}.

\begin{algorithm} 
\caption{Error-detection algorithm for process $q$} 
\label{fig.algo.detection}
\begin{algorithmic}
\Procedure{Error-Detection}{$S_q, S_{q-1}$}
  \State Set \textit{$error$} to false
  \State Send \textit{$data_q(S_{q-1})$} to process $q-1$
  \State Receive \textit{$data_{q+1}(S_{q})$} from process $q+1$
  \For{particle $p$ in \textit{$S_q$}}
  	\If {\textit{$data_{q+1}(p)$} differs from \textit{$data_q(p)$}}
  		\State Set \textit{$error$} to true
	\EndIf
  \EndFor
  \State \Return \textit{$error$}
\EndProcedure
\end{algorithmic}
\end{algorithm}

For a given process $q$, Algorithm~\ref{fig.algo.detection} iterates through the list of replicated particles that have been selected on process $q$ and computed on process $q+1$. For each selected particle, it compares the data obtained on process $q+1$, corresponding to the replica, with the data obtained on process $q$, corresponding to the original particle. If the data differ, an error has been detected.





\subsection{Limitations}

The SPR detection capabilities are limited to errors that propagate. SPH codes typically simulate several complex physical phenomena in the same application. Some particle datasets, such as the positions of the particles, are used very often during interpolation throughout the simulation. The data is read and used multiple times in different interpolation steps. An error in such datasets is more likely to propagate, and therefore to be detected, while errors in datasets than are used less often, such as particle particle densities, are less likely to propagate and may not always be detected.

Note that this approach is limited to deterministic calculations and that round-off errors are expected and need to be accounted for. This is typically achieved by allowing $k$ lower order bits to differ. The SPH method is entirely deterministic, and in the experiments performed in this work, no round-off errors or truncated values were encountered. 

\subsection{Error Correction}

Even though the main focus of this work is on the detection of SDCs, it is possible to combine the proposed detector with other error-detection and correction methods. 
Checkpointing with rollback recovery~\cite{CL85,uncoordinated} is the de-facto general-purpose recovery technique in high-performance computing.
Finding the optimal checkpointing interval~\cite{young74,daly04,Di16_Twolevel,benoit16_ipdps} or the optimal recovery method for SPH codes is beyond the scope of this paper.

Because the SPR error detector is used in every simulation time-step, it is possible to safely checkpoint the state of the simulation if no error was detected. 
Then, whenever a new error is detected, the simulation can simply rollback to the last correct checkpoint and re-execute from there.
For completeness, this work implements checkpointing using the Fault-Tolerance Interface (FTI)~\cite{6114441}.


\section{Experimental Evaluation}\label{sec:experiments}

In this section, we describe the experiments conducted with two test simulations commonly used in astrophysics. 
The goal of these experiments is two-fold: 
(1) experimentally assess the detection capabilities of SPR through an error injection campaign;
and (2) assess the performance of the proposed approach at scale.

\subsection{Experimental Setup}

\paragraph*{SPR implementation into a production SPH code} 
SPR has been incorporated in \mbox{SPHYNX}\footnote{Freely available at \url{https://astro.physik.unibas.ch/sphynx}}~\cite{cabezon2017}, an SPH code with focus on astrophysical simulations.
The incorporation of SPR into SPHYNX comprises the implementation of Algorithms~\ref{fig.algo.selection}, ~\ref{fig.algo.sph+gr}, and~\ref{fig.algo.detection} presented in Section~\ref{sec:SPR}. 
SPHYNX includes state-of-the-art SPH methods that allow it to address subsonic hydrodynamical instabilities and strong shocks, which are ubiquitous in astrophysical scenarios. SPHYNX is implemented in Fortran and uses the message-passing interface (MPI) for inter-process communication and the open multi-processing interface (OpenMP) for intra-process/inter-threads data sharing.
Note that compared to the original SPHYNX code, which counts $\sim 30,000$ lines of code, less than $300$ additional lines of code were added by SPR as a module for SPHYNX.
In the following, the term SPHYNX denotes the original code, and SPHYNX+SPR denotes the version of the code extended with SPR.
SPHYNX was used to run two different test simulations: 
(1)~the Evrard collapse (EC)~\cite{evrard1988}, which studies the gravitational collapse of a gaseous cloud and is a common test to evaluate the correctness of the coupling of hydrodynamics and self-gravity; and
(2)~the \mbox{wind-bubble} interaction test (WB)~\cite{agertz2007}, which studies the interaction of a supersonic wind with a high-density colder bubble in pressure equilibrium.

\paragraph*{Choice of number of neighbors}
The number of neighbors per particle must cover the area of influence of the interpolation kernel in a reasonably homogeneous way, and is, therefore, linked to the type of SPH kernel used in the simulation. 
In general, $10^{2}$ neighbors is a very common value found in the bibliography, for frequently used SPH kernels in 3D simulations.
Unlike the kernel used in SPHYNX (Sinc kernel \cite{cabezon2008}), other SPH kernels, e.g.,~Wendland kernel \cite{dehnen2012}, have a much longer radius and, as a consequence, must use larger amounts of neighbors (e.g., $5\times 10^{2}$).
These choices are made after an empirical iterative process in the SPH history and are proven to be adequate to balance accuracy and computational cost, while suppressing numerical instabilities, like particle pairing.

\paragraph*{Experiments}
For each test simulation, different parameters were passed to SPHYNX: 
a file containing the initial conditions, the corresponding number of particles, and other parameters specific to each test. 
During the time this work was performed, the initial conditions for only three tests cases were available: 
an EC test simulation with $65,536$ particles that was used to experiment with error-injection and detection, and two larger EC and WB test simulations with $1,000,000$ and $3,157,385$ particles, respectively, that were used to assess the scalability of SPR.
Generating initial conditions for different numbers of particles is a \mbox{\textit{non-trivial}} process.
Therefore, this work employs a set of \mbox{strong-scaling} experiments to assess the performance at scale for the proposed SPR method.
Table~\ref{fig.experiments} summarizes the different experiments and the corresponding parameters used. Error-injection and error-detection experiments were conducted on a small scale system with $20$ nodes named \emph{miniHPC}, while strong-scaling experiments were done the \emph{Piz Daint} supercomputer with up to $256$ nodes. 

\begin{table}
\centering
\begin{tabular}{ l | c | c | c }
\textbf{Experiment target} & \multicolumn{1}{c|}{\textbf{Error-injection}} & \multicolumn{2}{c}{\textbf{Strong-scaling}}\\
\hline
\hline
Test simulation & ECsg & EC & WB \\
\hline
\#Processes & 16 & 4-256 & 4-256\\
\hline
\#Threads/process & 20 & 12 & 12 \\
\hline
\#Particles & $65,536$ & $1,000,000$ & $3,157,385$ \\
\hline
Self-gravity & Yes & Yes & No  \\
\hline
HPC system & \multicolumn{1}{c|}{miniHPC} & \multicolumn{2}{c}{Piz Daint}\\
\hline
\#Neighbors & \multicolumn{3}{c}{100} \\
\end{tabular}
\caption{Design of target experiments}
\label{fig.experiments}
\end{table}

\subsection{Error-Injection and Error-Detection Experiments}
\label{sec:detection}

In this section, an error-injection and detection campaign was performed using the ECsg test simulation described in Table~\ref{fig.experiments}. 
The goal was to experimentally assess the detection capabilities of SPR. 
Specifically, the focus was on measuring the recall and the precision of the detector, where the precision is defined as $100 \times \left(1-\frac{\text{\#false-positives}}{\text{\#errors detected}}\right)$, and the recall is defined as $100 \times \frac{\text{\#errors detected}}{\text{\#errors detected} + \text{\#errors undetected}}$.

A \emph{golden set} of results was first created by running two time-steps of the application in an error-free environment.
To inject errors, an additional thread was created in each process to inject single bit-flips into the data of the application during its execution. 
The timing of the error to be injected is \emph{critical}: if it is injected too early in a time-step, the error may be overwritten and masked, while if it is injected too late (after propagation occurred), it may not be detected by SPR.

The simulation repeatedly executed the same two \mbox{time-steps} as in the golden set: a single bit-flip was injected at a random time and a random data location within the first \mbox{time-step}, and results were collected at the end of the second time-step.
The injected error was \emph{masked} if all datasets were identical to the ones in the golden set, \emph{detected} if SPR raised a flag, and \emph{undetected} otherwise. 
\emph{False-positive} errors are also counted, which would have been observed if SPR had raised a flag while there was no error. 
However, \emph{no false-positives were observed in the experiments}, which means that SPR has an experimentally measured precision of $100\%$. 

\begin{figure}[!htb]
	\centering
		\begin{subfigure}{6cm}
			\vspace{-0.25cm}
			\includegraphics[scale=0.465]{{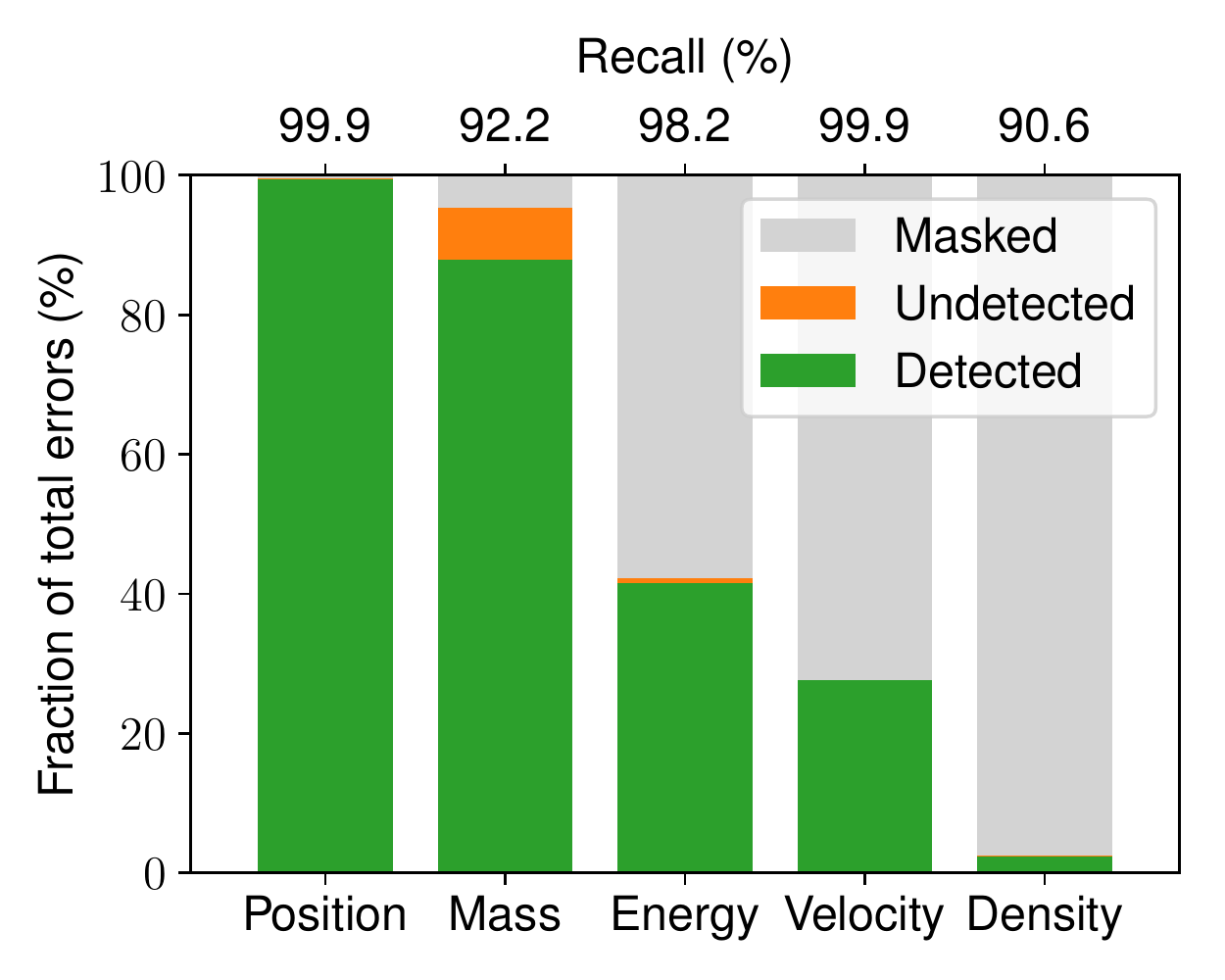}}
			\vspace{-0.25cm}
			\caption{Significant errors}\label{fig.recall-large}
		\end{subfigure}
		\begin{subfigure}{6cm}
			\includegraphics[scale=0.465]{{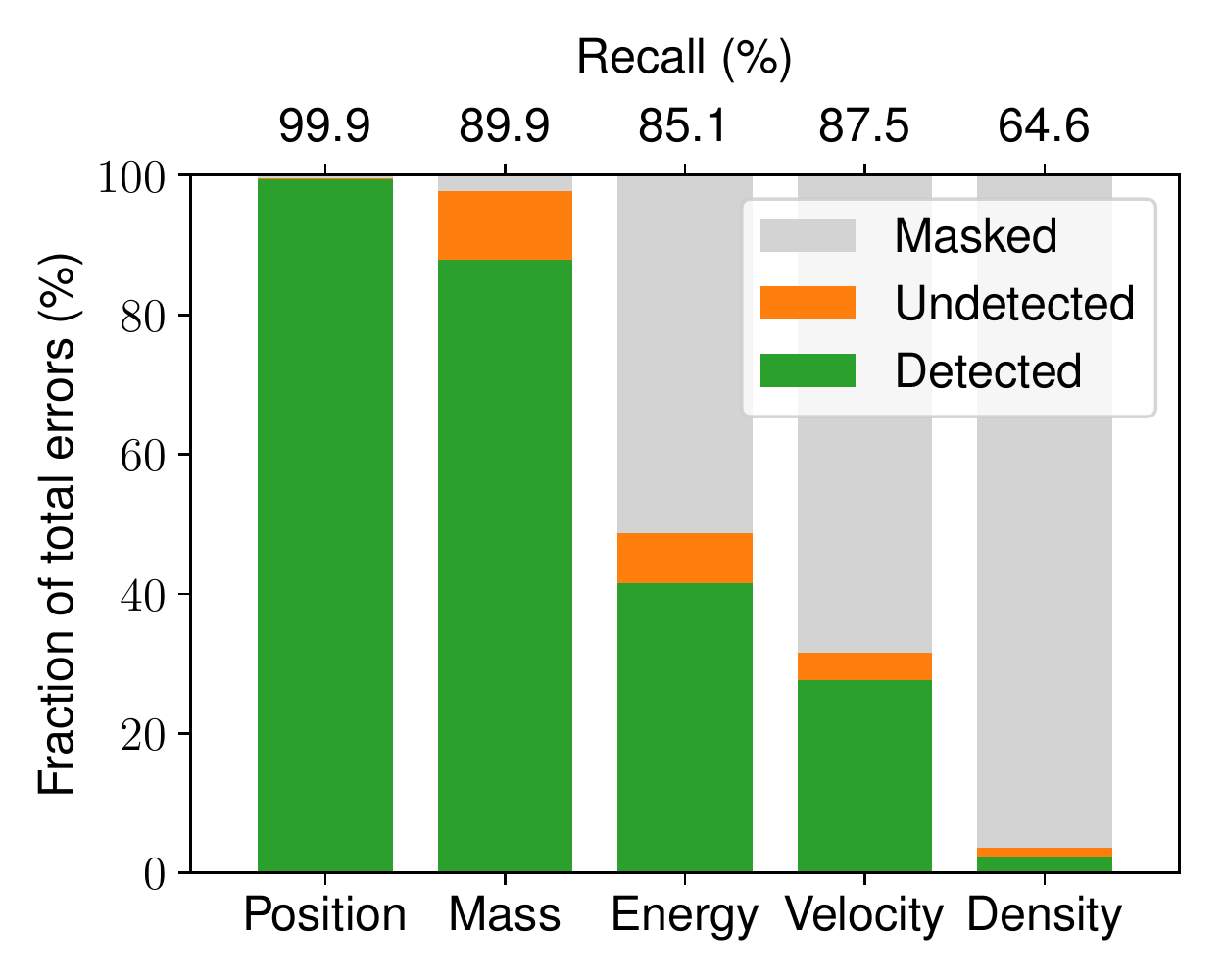}}
			\vspace{-0.25cm}
			\caption{All errors}\label{fig.recall-small}
		\end{subfigure}
	\caption{Fraction of errors detected, undetected, and masked for five particle datasets, where (a) only accounts for significant errors (in the sign, exponent, or first 5 digits), while (b) accounts for all errors (in any of the $16$ decimal digits). $25,000$ errors were injected over the execution of $50,000$ simulation time-steps.}
	\label{fig.detection}
\end{figure}

To quantify the recall, experiments were performed of the most five critical particle datasets, namely position, mass, internal energy, velocity and density. 
Each one of these datasets holds as many $64$-bits double-precision numbers as there are particles in the simulation, and up to three times more for multi-dimensional datasets, such as positions and velocities.
Following the IEEE 754 number representation,
bit $64$ encodes the sign, bits $63$ to $53$ (included) encode the exponent, and bits $53$ to $1$ encode the fraction or mantissa, which corresponds to approximately $16$ decimal digits of precision.
For each of these datasets, $5,000$ bit-flips were injected in random processes, particles, and bit positions, for a total of $25,000$ errors injected over the execution of $50,000$ simulation time-steps.

The results of the error-injection and detection campaign are presented in Figure~\ref{fig.detection}. 
Note that certain SDCs fall into the significant errors category. 
These errors correspond to an absolute difference greater than $10^{-6}$ between the corrupted and the golden data. 
Most detection techniques focus only on significant errors and will fail to detect less significant errors (of which many of them may be impactful). In fact, only a comparison with the actual correct data can reveal them, which SPR is able to perform.

The differences in the number of detected, masked, and undetected errors for different datasets are due to several factors. 
Certain particle datasets, such as positions, are critical to the SPH method. 
They are used in almost every computational step of the workflow, which greatly increases the probability that an error is propagated, and therefore detected by SPR. 
Other particle datasets, such as density, can be recomputed from other datasets, and are less frequently used in the computational workflow.
Therefore, the probability that an error is masked is higher, and there are fewer opportunities for SPR to detect it.

\begin{figure}[!htb]
	\centering
		\begin{subfigure}{6cm}
		\centering\captionsetup{width=1.2\linewidth}%
			\vspace{-0.2cm}
			\includegraphics[scale=0.55]{{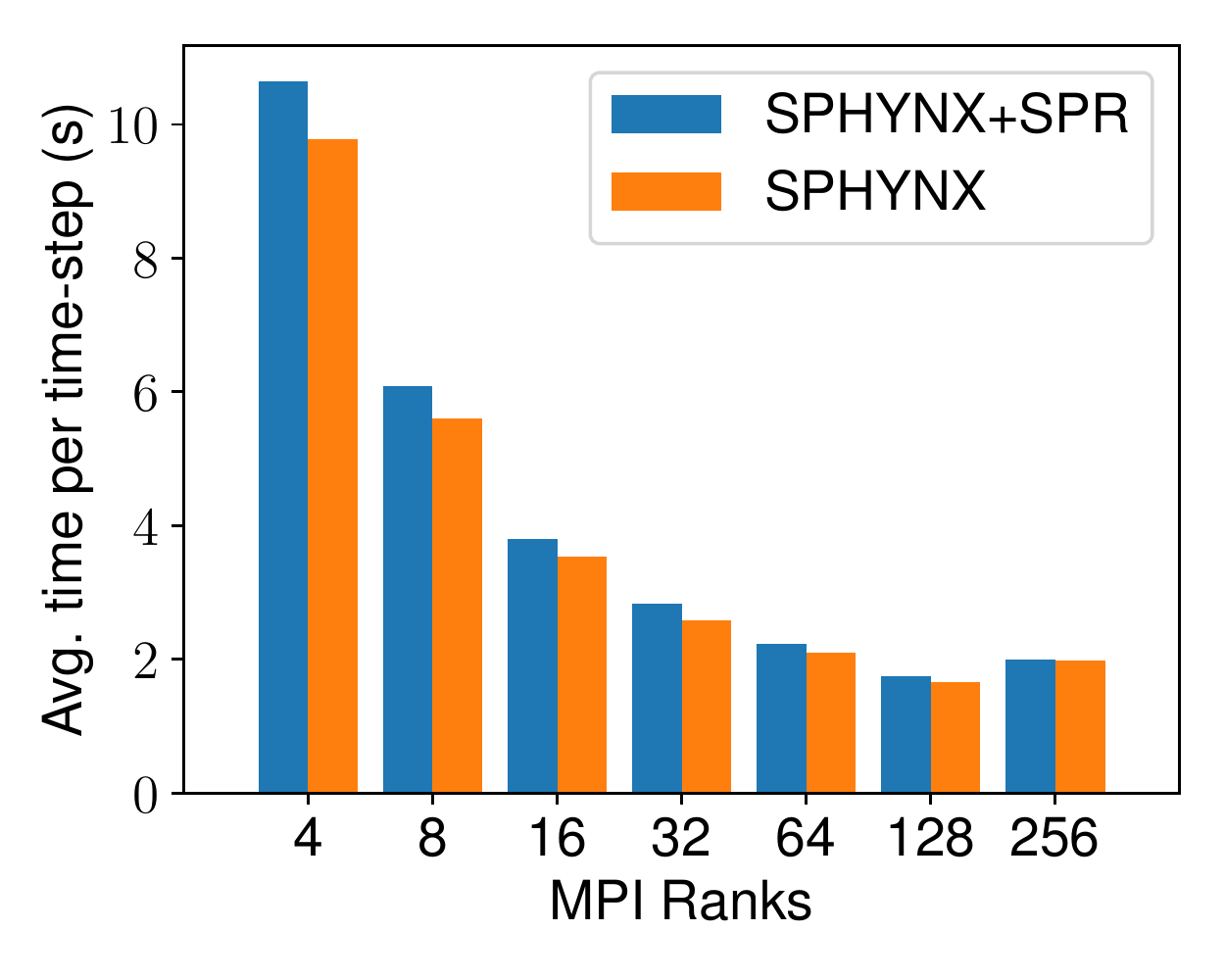}}%
			\vspace{-0.3cm}
			\caption{Average execution time per time-step (EC)}\label{fig.time-evrard}
		\end{subfigure}
		\begin{subfigure}{6cm}
		\centering\captionsetup{width=1.2\linewidth}%
			\includegraphics[scale=0.55]{{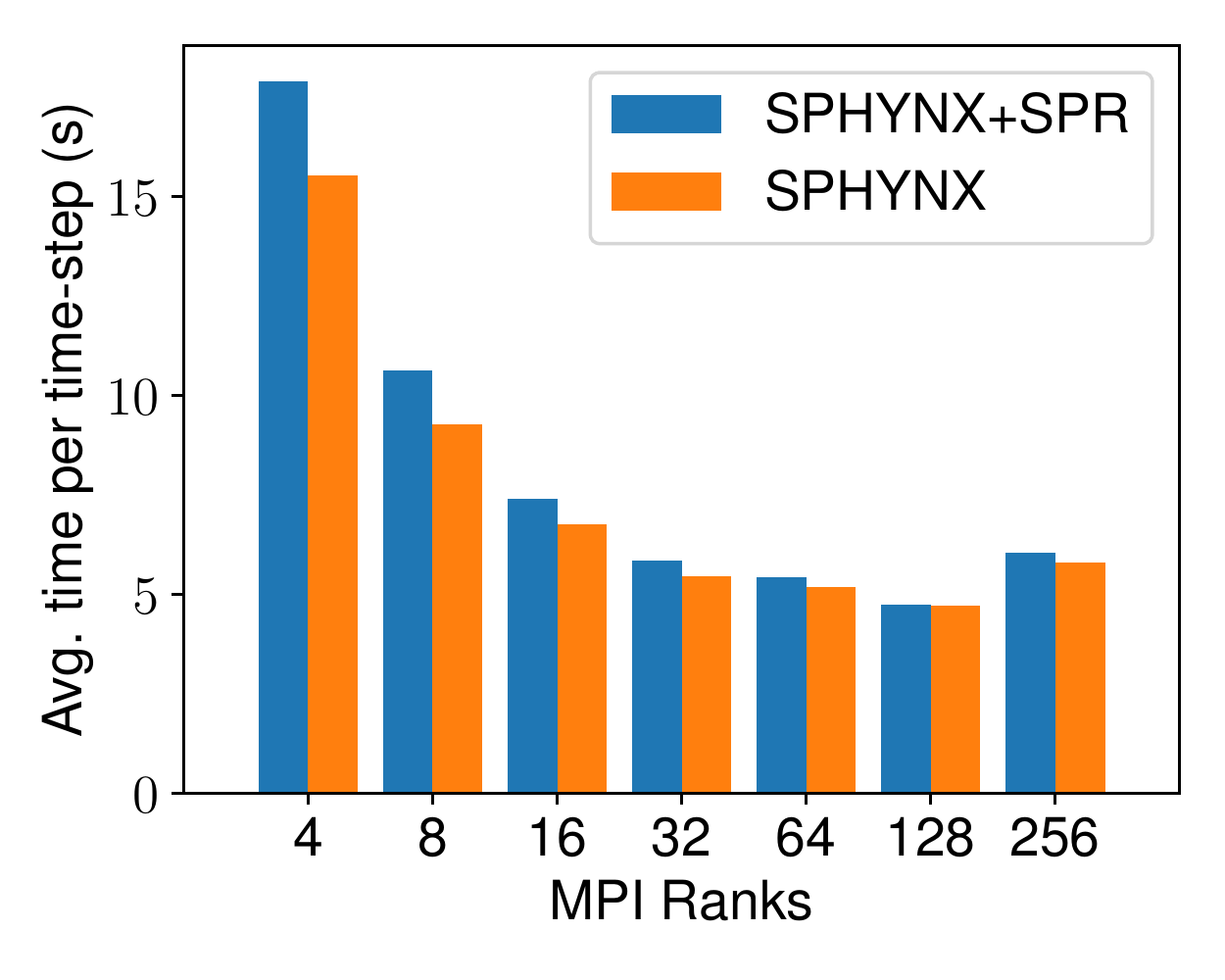}}
			\vspace{-0.75cm}
			\caption{Average execution time per time-step (WB)}\label{fig.time-windblob}
		\end{subfigure}
	\caption{Average time needed to complete one time-step for the EC test with the resilient code (SPHYNX+SPR) and the original version (SPHYNX) for the EC test (a) and the WB test (b).}
	\label{fig.evrard-scaling}
\end{figure}
\subsection{Strong Scaling Experiments}
\label{sec:strong_scaling}

Overall, the measured recall was always greater than $0.91$, and as high as $0.999$ for significant errors. 
The recall is only slightly lower when accounting for all errors below the current precision limit. 

\begin{figure}[!htb]
	\centering
		\begin{subfigure}{6cm}
		\centering\captionsetup{width=1.2\linewidth}%
			\vspace{-0.2cm}
			\includegraphics[scale=0.55]{{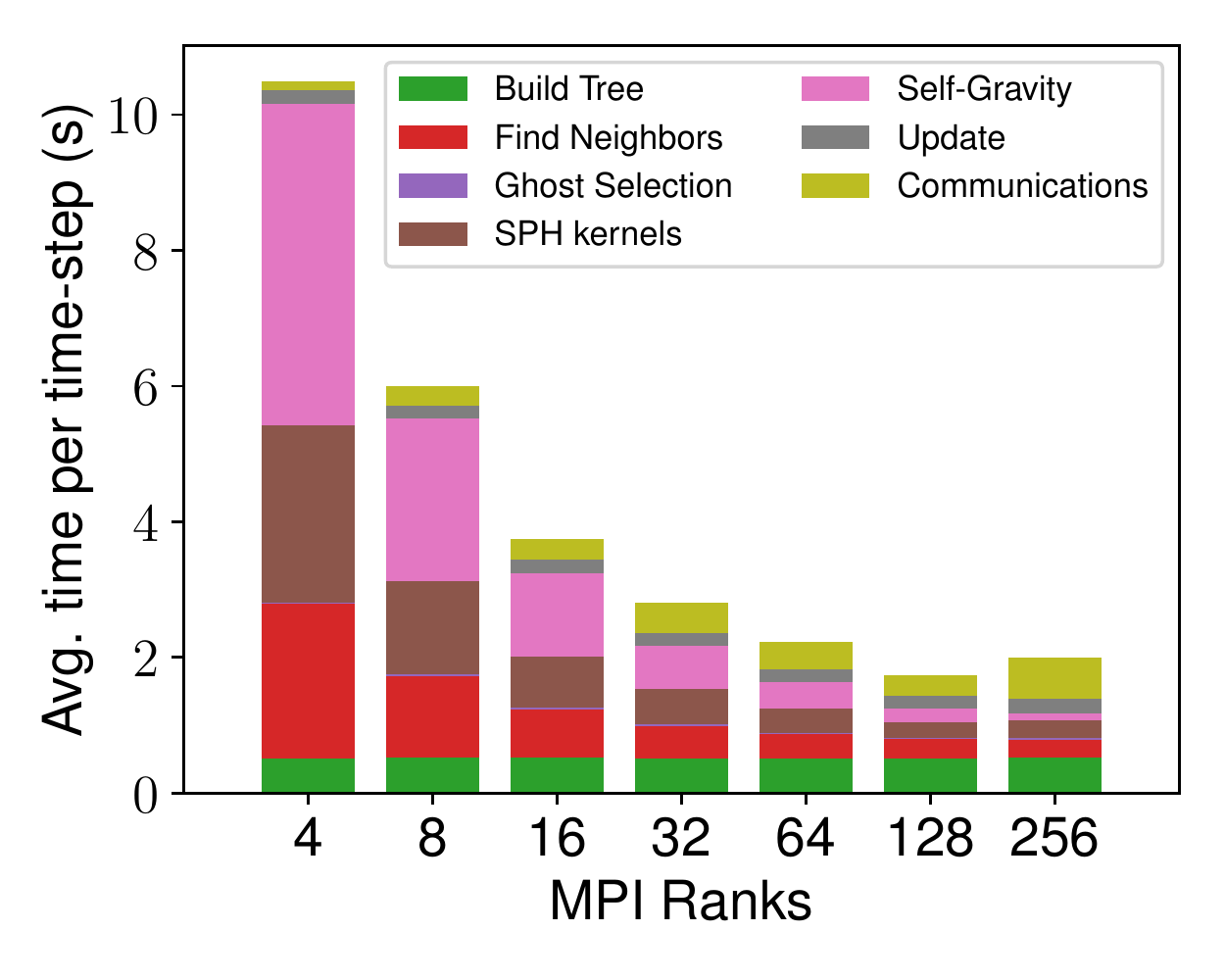}}
			\vspace{-0.75cm}
			\caption{Execution break down for SPHYNX+SPR by computational workflow step within a single time-step (EC)}\label{fig.trace-evrard}
		\end{subfigure}
		\begin{subfigure}{6cm}
		\centering\captionsetup{width=1.2\linewidth}%
			\includegraphics[scale=0.55]{{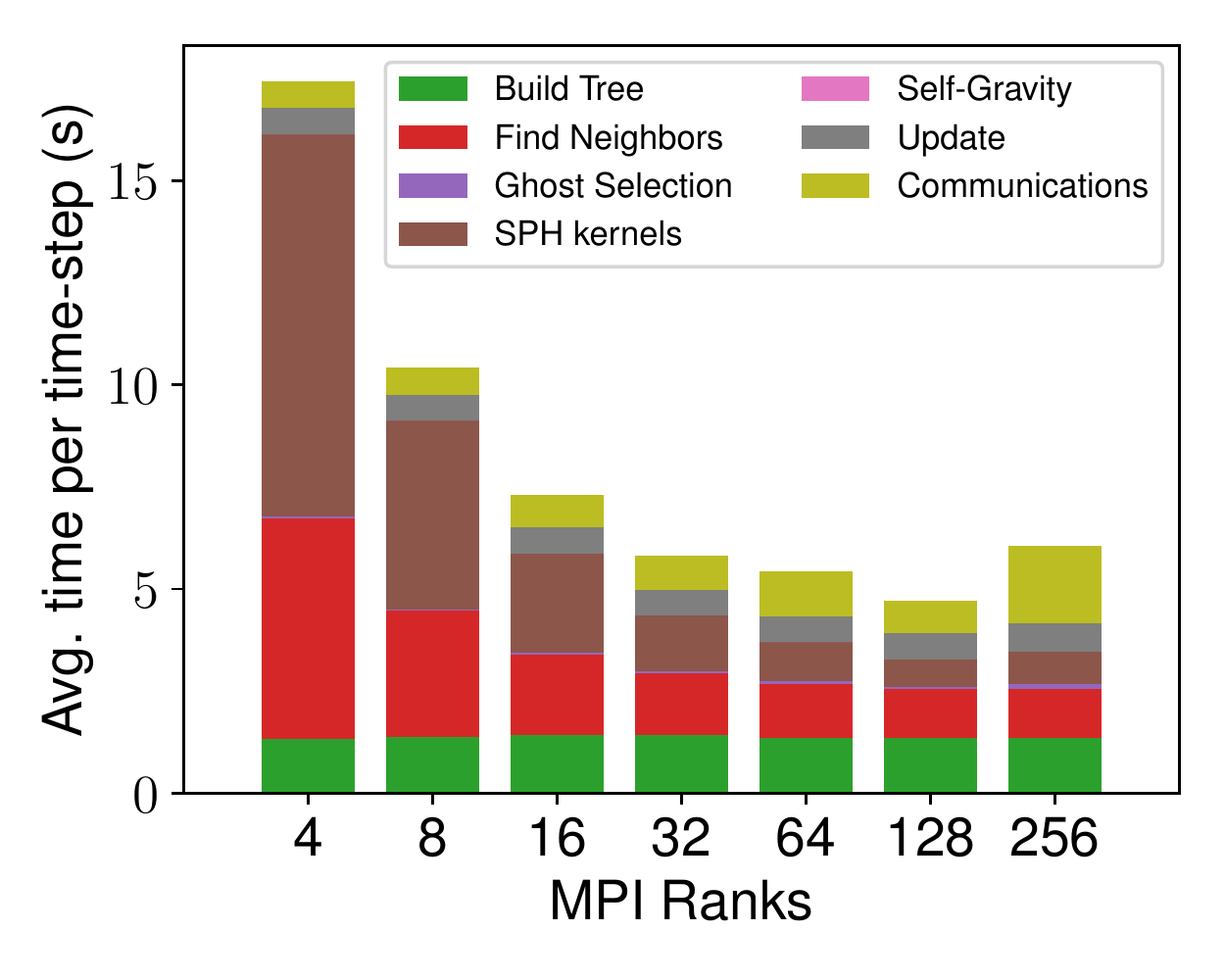}}
			\vspace{-0.75cm}
			\caption{Execution break down for SPHYNX+SPR by computational workflow step within a single time-step (WB)}\label{fig.trace-windblob}
		\end{subfigure}
	\caption{Break down by workflow step of the time needed to complete one time-step for the EC test (a) and WB test (b) with the resilient code (SPHYNX+SPR) and the original version (SPHYNX).}
	\label{fig.windblob-scaling}
	\vspace{-0.5cm}
\end{figure}

To assess the performance of SPR at scale, strong-scaling experiments were conducted on the EC and WB test cases, as described in Table~\ref{fig.experiments}.

Each experiment was performed by running two versions of SPHYNX, the original version (SPHYNX) and the resilient version (SPHYNX+SPR).
As the focus of this experiment is on assessing the performance of error-detection at scale, given the long execution time requirements for such simulations, the number of time-steps of the simulation was set to $20$, while a full SPH simulation would typically require several thousands time-steps. 
For each execution, we have excluded the first $10$ time-steps that correspond to the SPH initialization phase and are not representative of the general performance behavior of the SPH code.
Overall, the reported execution time of each test case is an average of a total of $100$ time-steps for $10$ executions.

Figures~\ref{fig.time-evrard} and~\ref{fig.time-windblob} present the average execution time needed to complete a single time-step for both SPHYNX and SPHYNX+SPR, for the EC and WB test cases, respectively, when executed on $4$ to $256$ computing nodes, each with $12$ computing cores.
In both cases, it can be observed that SPHYNX+SPR yields a low overhead, never exceeding $5\%$ and $10\%$ for the EC and WB test cases, respectively. 
This corresponds to the case with the largest number of particles per node.

Figures~\ref{fig.trace-evrard} and~\ref{fig.trace-windblob} illustrate a break down of the execution time of \mbox{SPHYNX+SPR} by the main computational workflow steps, including the time to select particles and the total communication time.
It can be seen that the algorithm to select particles represents at most $2.6\%$ of the time needed to complete a single time-step.
The total cost of detection never exceeds $0.1\%$ of the total execution time and is not shown in these figures.

Overall, both SPHYNX and \mbox{SPHYNX+SPR} scale very well with increasing node count, the main performance limitations being: 
(1)~the sequential implementation of the Build tree step SPHYNX; and 
(2)~the increasing communication cost, which is negligible when executing on $4$ computing nodes, but a dominant factor when executing on more than $128$ nodes.

\begin{figure}[!htb]
	\centering
		\begin{subfigure}{6cm}
		\centering\captionsetup{width=1.2\linewidth}%
			\vspace{-0.2cm}
			\includegraphics[scale=0.55]{{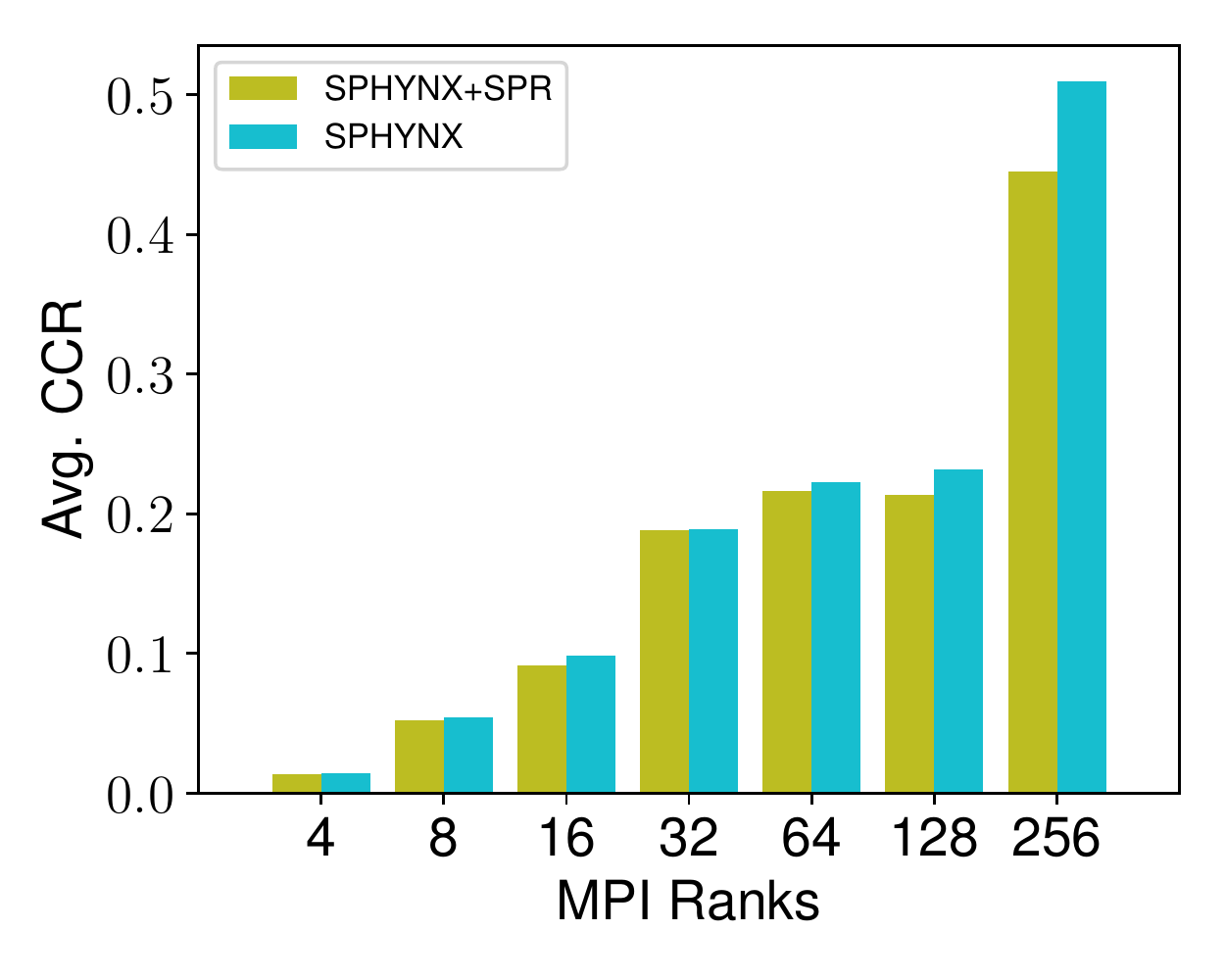}}
			\vspace{-0.75cm}
			\caption{Average CCR (EC)}\label{fig.comms-evrard}
		\end{subfigure}
		\begin{subfigure}{6cm}
		\centering\captionsetup{width=1.2\linewidth}%
			\includegraphics[scale=0.55]{{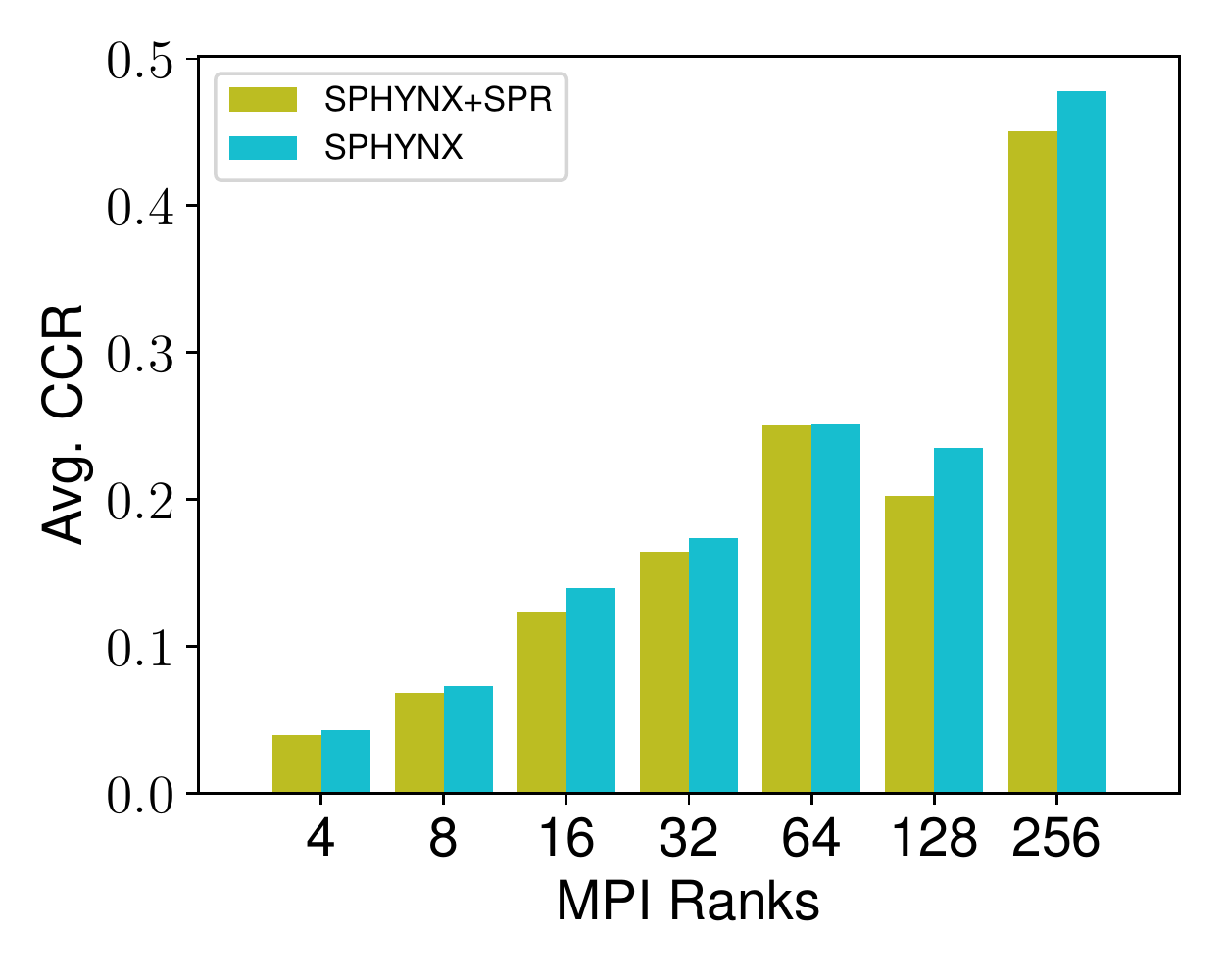}}
			\vspace{-0.75cm}
			\caption{Average CCR (WB)}\label{fig.comms-windblob}
		\end{subfigure}
	\caption{Average communication to computation ratio (CCR) per time-step for the EC test (a) and the WB test (b) with the resilient code (SPHYNX+SPR) and the original version (SPHYNX).}
	\label{fig.comms}
\end{figure}

Figures~\ref{fig.comms-evrard} and Figure~\ref{fig.comms-windblob} show the average communication to computation ratio (CCR) for the EC and WB test simulations, respectively. For both versions of the code, the CCR ratio steadily increases with the number of nodes. Yet, despite requiring multiple additional inter-process communications, the difference with between SPHYNX and SPHYNX+SPR remains very small.
The significant increase in overhead on $256$ nodes (also shown in Figure~\ref{fig.trace-evrard} and Figure~\ref{fig.trace-windblob}) is mainly due to having too few particles per node.

\subsection{Summary and Discussion}

The experimental results show that SPR can practically be incorporated into commonly used astrophysical SPH test simulations with minimal changes to the original code, and with low additional overhead, ranging from $1\%$ to $10\%$ for the EC and WB test cases.

SPH simulations are expected to use more particles in production runs (i.e. up to $10^{12}$), which results in greater particle counts per process and greater neighbor count per process (i.e. several hundreds) in some scenarios. 
Consequently, 
SPR is expected to deliver improved performance on \mbox{large-scale} SPH simulations: an increase in both the number of particles and number of neighbors per particles will lead to fewer selected particles, and, thus, less overhead overall.
%


\section{Related Work}\label{sec:relwork}

\paragraph*{Complete redundancy}

The use of redundant MPI processes for error detection has been widely analyzed in the last decade~\cite{j137,MR-MPI,Ferreira2011,Fiala12Detection,thread-rep1,thread-rep2,thread-rep3}. 
Unlike these full-process or full-task replication efforts, the present work employs selective \mbox{\emph{sub-process}} or \mbox{\emph{sub-task}} duplication (illustrated in Figure~\ref{fig.SPR-replication}) of a selected part of the work assigned to a process or a task that corresponds to the particle selected for replication. 
While dual modular redundancy or triple modular redundancy incurs $100\%$ and $200\%$ additional computational overhead, respectively, SPR only replicates $1-10\%$ of the application with a $1-10\%$ added overhead.

\paragraph*{Partial redundancy}

Partial redundancy has been studied to decrease the overhead of complete redundancy~\cite{Elliott2012,ftxs12-rep,Subasi2015,Subasi2017}.
Adaptive partial redundancy has also been proposed wherein a subset of processes is dynamically selected for replication~\cite{George2012}.
Partial replication (using additional hardware) of selected MPI processes 
has been combined with \mbox{prediction-based} detection to achieve SDC protection levels comparable with those of full duplication~\cite{Berrocal16Exploring,Berrocal17Toward,FlipBack16}.
The proposed SPR approach differs from the existing partial redundancy approaches in that it protects the data of the entire application (as opposed to a subset) by selectively duplicating \emph{only a subset} of the computations \emph{within} processes. 
To the best of our knowledge, this is the first time such methods are applied to SPH.

\paragraph*{Detection via Interpolation}

Silent error detectors based on \emph{data analytics} use several interpolation techniques, such as time series prediction~\cite{Berrocal15Lightweight} 
and multivariate interpolation~\cite{PPoPP14,Gomez15Detecting,Gomez15Exploiting}, to \emph{interpolate} the next value of a computational point based on spatial and temporal data smoothness.
Multivariate interpolation has been used to detect and correct SDCs in computational fluid dynamics~\cite{Gomez15Detecting}, where the approach achieves $99.8\%$ precision and low recall.
Additional tests on synthetic benchmarks achieved a maximum of $90\%$ recall at $1\%$ overhead~\cite{Gomez15Exploiting}.
To ensure that the false-positive rate remains well below the SDC rate, a high detection precision supersedes a high recall.
As shown in Section~\ref{sec:experiments}, SPR yields $100\%$ precision (i.e. no false-positives) and a high recall for selected data-fields.

\emph{Algorithm-based fault~tolerance} (ABFT) is another error detection technique which uses checksums to detect up to a certain number of errors and is currently only suitable for linear algebra kernels~\cite{Kuang1984,bosilca2009algorithm,Shantharam2012}.

\section{Conclusion and Future Work}\label{sec:conclusion}

In this work, we proposed a novel silent data corruption (SDC) detection method, namely selective particle replication (SPR), and we implemented it in SPHYNX, a production smoothed particle hydrodynamics (SPH) simulation code.

SPR comprises of three simple, yet efficient, algorithms to select the particles for which to replicate the computations and data of selected particles onto different processes, and to compare the values of the original particles against the data of their replica, to detect SDCs during execution.

The error-detection capabilities of SPR were experimentally evaluated through an error-injection and detection campaign. 
Not only is SPR able to detect $91-99.9\%$ of the injected errors (i.e., DRAM bit-flips), it also does not raise any false-positives.
In addition, experiments conducted on an HPC system demonstrate the scalability of SPR in production SPH codes, at an overhead of $1-10\%$.

Because SPR is scalable, non-intrusive and precise, it can easily be combined with other error-detection methods, to detect errors that could escape SPR's detection coverage.
Furthermore, SPR can also be applied to other classes of applications, e.g., \mbox{N-body} simulations, stencils, and computational fluid dynamics.
Whether SPR can accurately detect errors in such applications is a topic that deserves further investigation.

\section*{Acknowledgement}

This work has been supported by the Swiss Platform for Advanced Scientific Computing (PASC) project SPH-EXA and by a grant from the Swiss National Supercomputing Centre (CSCS) under project ID c16.

\bibliographystyle{abbrv}
\bibliography{biblio}

\end{document}